# Image-based Reconstruction for a 3D-PFHS Heat Transfer Problem by ReConNN

**Yu Li**[a, *], **Hu Wang**[a, **], **Xinjian Deng**[a]

*a. State Key Laboratory of Advanced Design and Manufacturing for Vehicle Body, Hunan University, Changsha, 410082, PR China*

**Abstract**

The heat transfer performance of Plate Fin Heat Sink (PFHS) has been investigated experimentally and extensively. Commonly, the objective function of the PFHS design is based on the responses of simulations. Compared with existing studies, the purpose of this study is to transfer from analysis-based model to image-based one for heat sink designs. Compared with the popular objective function based on maximum, mean, variance values etc., more information should be involved in image-based and thus a more objective model should be constructed. It means that the sequential optimization should be based on images instead of responses and more reasonable solutions should be obtained. Therefore, an image-based reconstruction model of a heat transfer process for a 3D-PFHS is established. Unlike image recognition, such procedure cannot be implemented by existing recognition algorithms (e.g. Convolutional Neural Network) directly. Therefore, a Reconstructive Neural Network (ReConNN), integrated supervised learning and unsupervised learning techniques, is suggested and improved to achieve higher accuracy. According to the experimental results, the heat transfer process can be observed more detailed and clearly, and the reconstructed results are meaningful for the further optimizations.

**Nomenclature -- variable**

| | | | |
|---|---|---|---|
| *L* | length of the heat sink | *r* | second moment estimation |

[*] First author. *E-mail address*: liyu_hnu@hnu.edu.cn (Y. Li)

[**] Corresponding author. Tel.: +86 0731 88655012; fax: +86 0731 88822051.
  *E-mail address:* wanghu@hnu.edu.cn (H. Wang)

| | | | |
|---|---|---|---|
| $W$ | width of the heat sink | $\hat{y}_i$ | predicted value |
| $H$ | height of the heat sink | STD | standard deviation |
| $L_h$ | length of the heat source | MSE | mean square error |
| $W_h$ | width of the heat source | **x** | sample set |
| $t_0$ | the thickness of the baseplate | $p(y/\mathbf{x})$ | softmax output of label |
| $t_1$ | the thickness of each pin | $p(y)$ | overall labels distribution |
| $S$ | the distance between neighbor fins | | |
| $n$ | the number of fins | *Greek symbols* | |
| $t_w$ | temperature of wall | $\tau$ | $\tau$ |
| $t_f$ | temperature of fluid | $\lambda$ | $\lambda$ |
| $n$ | exterior normal of wall | $\rho$ | $\rho$ |
| $h$ | coefficient of heat transfer | $\Delta$ | $\Delta$ |
| $c$ | specific heat capacity | $\eta$ | $\eta$ |
| $\dot{\Phi}$ | internal heat source | $\lambda$ | $\lambda$ |
| $u$ | velocity component along $x$-axis | $\theta$ | $\theta$ |
| $v$ | velocity component along $y$-axis | $\rho_1$ | $\rho_1$ |
| $w$ | velocity component along $z$-axis | $\rho_2$ | $\rho_2$ |
| $F_x$ | body force along $x$-axis | $\varepsilon$ | $\varepsilon$ |
| $F_y$ | body force along $y$-axis | $\mu$ | $\mu$ |
| $F_z$ | body force along $z$-axis | $\mathbb{E}$ | operation of mean |
| $p$ | pressure | | |
| $c_p$ | specific heat at constant pressure | *Subscripts* | |
| $L(x)$ | loss function | $w$ | $w$ |
| $f(x)$ | the predicted value | $f$ | $f$ |
| $y_i$ | the real value | $x$ | $x$ |
| $x_i$ | training sample | $y$ | $y$ |
| $m$ | the number of $x_i$ | $z$ | $z$ |
| $s$ | first moment estimation | $p$ | $p$ |

**Nomenclature -- abbreviation**

| | | | |
|---|---|---|---|
| 2D | Two-Dimensional | MSE | Mean Square Error |
| 3D | Three-Dimensional | NHT | Numerical Heat Transfer |
| AdaGrad | Adaptive Gradient | NN | Neural Network |
| Adam | Adaptive Moment Estimation | PFHS | Plate Fin Heat Sink |
| CAD | Computer Aided Design | PPF | Parallel-Plain Fin |
| CA[*] | Convolutional Autoencoder | $R^2$ | $R$ Square |
| CIC | Convolution in Convolution | RAAE | Relative Average Absolute Error |
| CNN | Convolutional Neural Network | RE | Relative Error |
| CPU | Central Processing Unit | ReConNN | Reconstructive Neural Network |
| CWGAN | Compressed WGAN | ReLU | Rectified Linear Unit |
| DL | Deep Learning | RMAE | Relative Maximum Absolute Error |

---

[*] In order to differ with Computer Aided Engineering (CAE), the Convolutional Autoencoder is shorted as CA.

| | | | |
|---|---|---|---|
| GAN | Generative Adversarial Network | RMSProp | Root Mean Square Prop |
| IPFM | Interleaved Parallelogram Fin Module | SIMP | Solid Isotropic Material with Penalization |
| IHCP | Inverse Heat Conduction Problem | STD | Standard Deviation |
| KL | Kullback-Leibler | VAE | Variational Autoencoder |
| LI | Lagrange Interpolation | VG | Vortex Generator |
| LReLU | Leaky ReLU | WGAN | Wasserstein GAN |
| ML | Machine Learning | | |

## 1. Introduction

Over recent years, the electronics industry is increasingly prosperous, and the increasing the heat transfer efficiency of electronic components has become an important topic of research and development. Wang [1] presented flow visualization and frictional results of enlarged fin-and-tube heat exchangers with and without the presence of vortex generators. He found that the delta winglet caused more intensely vortical motion and flow unsteadiness than the annular winglet. Then he [2] performed a 3D turbulent flow numerical simulation to improve heat transfer characteristics of wavy fin-and-tube heat exchangers. Ahmed [3] gave an overview about the early studies done in order to improve the performance of thermal systems. Additionally, he also summarized the recent experimental and numerical developments on the Numerical Heat Transfer (NHT). Huisseune [4] numerically investigated the effect of punching delta winglet vortex generators into the louvered fin surface near the wake region of each tube. Additionally, he [5] studied the influence of the louver and delta winglet geometry on the thermal and hydraulic performance of a compound heat exchanger. Sinha [6] performed numerical investigations pertaining to heat transfer enhancement of a plate-fin heat exchanger by using two rows of winglet type Vortex Generators (VG). Chiang [7] applied Taguchi method to predict and optimize the cooling performance of Parallel-Plain Fin (PPF) heat sink. In his study, the optimum design parameters of the lowest value of the highest temperature were found. Chingulpitak [8] summarized the publications with respect to research on the fluid flow directions and behaviors through heat sinks which offered guidelines for future researches. Chen [9] proposed a novel air-cooled heat sink profile termed as Interleaved Parallelogram Fin Module (IPFM). The IPFM not only gained the advantage of lower pressure drop, but also obtained a material saving. Moreover, he [10] studied a quick weight saving methodology with trapezoidal base heat sink applicable for electronic cooling application. Damook [11] investigated the benefits of using pin fin heat sinks with multiple perforations. In addition, an experimental heat sink with multiple perforations was also designed. Wang

As air flows over a network of fins, convective heat transfer is the most common technique to cool electronics for low cost, availability and reliability [12]. The surface fins are able to achieve large heat transfer area and act as turbulence promoters for a further enhance of heat transfer rate, without excessive primary surface area. PFHSs are commonly used due to their simple structure and ease of manufacturing. Many works about PFHSs have been studied, e.g. Joo [13] compared the thermal performance of optimized plate-fin and pin-fin heat sinks with a vertically oriented base plate, and proposed a new correlation of the heat transfer coefficient. Because of the many design parameters and the critical issue of developing both cost and thermal effective, Ventola [14] presented a novel thermal model of PFHS. Hossain [15] developed an analytical model to predict air flow and pressure drop. The model applied conservation of mass and momentum over the bypass regions and established flow channels between fins. Li [16] assessed the performance of PFHSs in a cross flow. Experimental results indicated that increasing the Reynolds number could reduce the thermal resistance of the heat sink. Subsequently, he [17] investigated the thermal-fluid characteristics of a flat-fin heat sink with a pair of vortex generators. Srisomporn [18] demonstrated the practical multiobjective optimization of PFHSs and pointed out the superiority of using a combined response surface method and multiobjective evolutionary optimizer. Zhou [19] proposed a multiparameter constrained optimization procedure to design the PFHSs by minimizing their rates of entropy generation and three cases demonstrated the feasibility of the algorithm. Rao [20] presented the multiobjective design optimization of a PFHS equipped with flow-through and impingement-flow air cooling systems. Chen [21] developed a multiobjective real-coded genetic algorithm for the optimal heat sink design problem. Chen [22] applied the inverse method in conjunction with the experimental temperature data to investigate the accuracy of the heat transfer coefficient on the fin in the PFHS. Kim [23] optimized thermal performance of a vertical PFHS under natural convection.

It is well known that some Machine Learning (ML) methods, including the Neural Network (NN) which is the core technology of Deep Learning (DL), have been utilized in NHT, such as Shiguemori [24] described a methodology for using NNs in an inverse heat conduction problem to determine the initial temperature profile on a slab with adiabatic boundary condition. Sablani [25] accomplished the Inverse Heat Conduction Problem (IHCP) dealing with the estimation of the heat transfer coefficient for a solid/fluid by using a NN. Baby [26] determined the time to reach a set-point temperature for aluminum finned heat sinks filled with the phase change material n-eicosane. Then he integrated the results with a NN to predict operating times. Czél [27] proposed a NN-based solution of the inverse heat conduction problem of identifying the temperature-dependent volumetric heat capacity function of a solid material. Razavi [28] used NN to model the thermal performance of an underground cold-water cistern. Aminian [29] applied NNs to predict the thermal conductivity of nanofluids. Colorado [30] employed NNs to design a physical–empirical model to describe heat transfer of helical coil in oil and glycerol/water solution.

In this study, the NN is applied to construct surrogate models for a PFHS to refine the heat transfer process for further optimizations. This NN-based reconstruction transfers from analysis-based model to image-based one. In other words, the structural optimizations should be based on images stead of responses. To complete this study, Reconstructive Neural Network (ReConNN) is employed. The ReConNN was proposed in Ref. [31], where it was originally applied to a simply 2-dimensional model based on Solid Isotropic Material with Penalization (SIMP). In this study, the method is improved to have a better result. At the same time, it is used to reconstruct for a more complex 3-dimensional model of a PFHS. For this purpose, a novel reconstruction method, 3D-2D-3D, for 3D models is designed. The slicing helps us to achieve the process of 3D-2D-3D simply. Ultimately, the reconstruction of a heat transfer process of a 3D-PFHS is studied to validate the effectiveness reconstruction method.

## 2. PFHS descriptions

### 2.1. Physical model

The 3D-steady-state Computer Aided Design (CAD) model of the PFHS is presented in Fig. 1. The heat sink is made of aluminum alloy material whose density is $2.702 \times 10^3 \text{kg/m}^3$. Furthermore, effects of gravity and radiative heat transfer are neglected. $L$, $W$ and $H$ are the length, width and height of the heat sink, respectively. $L_h$ and $W_h$ are the length and width of the heat source, respectively. $t_0$ is the thickness of the baseplate, and the thickness of each pin is presented by $t_1$. The distance between neighbor fins is $S$.

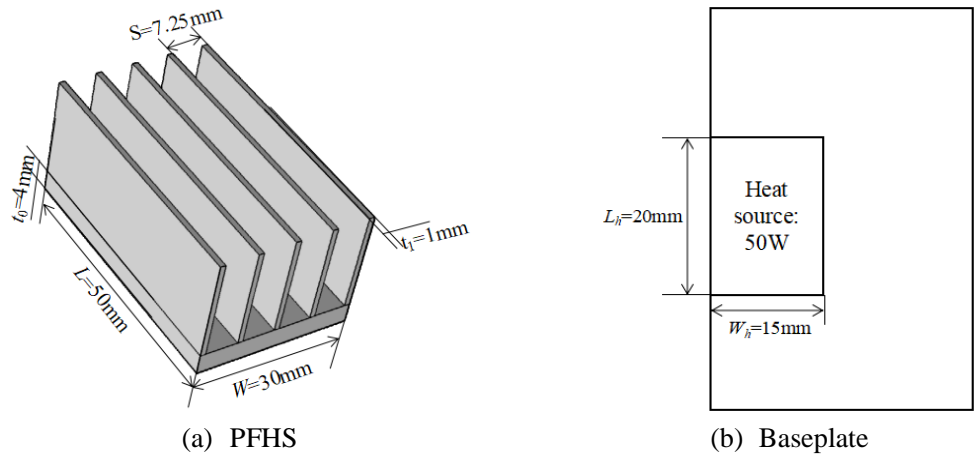

(a) PFHS    (b) Baseplate

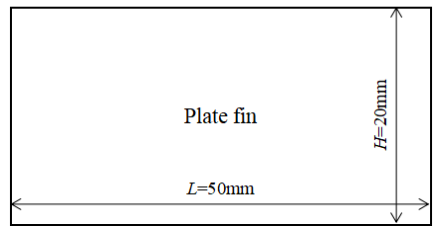

(c) Plate fin

**Fig. 1.** The CAD model of the PFHS.

The total volume [32] of fin volume is fixed.

$$V = LWH = \text{Constant} \tag{1}$$

The volume of each plate fin is

$$V_1 = LWt_1 \tag{2}$$

The number of fins is

$$n = \frac{W - t_1}{S} \tag{3}$$

In this study, the geometric parameters of the PFHS are marked in Fig. 1, and the number of PFHS is selected as 5. The boundary condition of the PFHS is a Robin problem.

*2.2. Mathematical model*

The differential equation of a 3D heat conduction in Cartesian coordinates is

$$\rho c \frac{\partial t}{\partial \tau} = \frac{\partial}{\partial x}\left(\lambda \frac{\partial t}{\partial x}\right) + \frac{\partial}{\partial y}\left(\lambda \frac{\partial t}{\partial y}\right) + \frac{\partial}{\partial z}\left(\lambda \frac{\partial t}{\partial z}\right) + \dot{\Phi} \tag{4}$$

where $\rho$ is density; $c$ is specific heat capacity; $\tau$ is time; $t$ is the temperature field; $\lambda$ is the thermal conductivity; and $\dot{\Phi}$ is internal heat source.

Considering the problem in this study is steady state, constant property, and having heat source, the differential equation can be simplified as a Poisson equation:

$$\frac{\partial^2 t}{\partial x^2} + \frac{\partial^2 t}{\partial y^2} + \frac{\partial^2 t}{\partial z^2} + \frac{\dot{\Phi}}{\lambda} = 0 \tag{5}$$

that

$$\Delta t + \frac{\dot{\Phi}}{\lambda} = 0 \tag{6}$$

where $\Delta$ is Laplace operator.

Commonly, the heat boundary conditions are:

- Dirichlet boundary condition: $t_w = f(\tau)$

- Neumann boundary condition: $-\lambda \left(\frac{\partial t}{\partial n}\right)_w = f(\tau)$

- Robin boundary condition: $-\lambda\left(\dfrac{\partial t}{\partial n}\right)_w = h(t_w - t_f)$

- Adiabatic boundary condition: $-\lambda\dfrac{\partial t}{\partial n} = 0$

where $t_w$ and $t_f$ are the wall temperature and environment temperature, respectively; $n$ is the unit outward normal to the boundary; and $h$ is the coefficient of heat transfer.

Governing equation can be expressed as follows.

Mass conservation equation:

$$\frac{\partial u}{\partial x} + \frac{\partial v}{\partial y} + \frac{\partial w}{\partial z} = 0 \tag{7}$$

where $u$, $v$ and $w$ are velocity components of fluid along $x$-axis, $y$-axis and $z$-axis, respectively.

Momentum conservation equation:

$$\begin{cases} \rho\left(\dfrac{\partial u}{\partial \tau} + u\dfrac{\partial u}{\partial x} + v\dfrac{\partial u}{\partial y} + w\dfrac{\partial u}{\partial z}\right) = F_x - \dfrac{\partial p}{\partial x} + \eta\Delta u \\ \rho\left(\dfrac{\partial v}{\partial \tau} + u\dfrac{\partial v}{\partial x} + v\dfrac{\partial v}{\partial y} + w\dfrac{\partial v}{\partial z}\right) = F_y - \dfrac{\partial p}{\partial y} + \eta\Delta v \\ \rho\left(\dfrac{\partial w}{\partial \tau} + u\dfrac{\partial w}{\partial x} + v\dfrac{\partial w}{\partial y} + w\dfrac{\partial w}{\partial z}\right) = F_z - \dfrac{\partial p}{\partial z} + \eta\Delta w \end{cases} \tag{8}$$

where $F_x$, $F_y$ and $F_z$ are body forces along $x$-axis, $y$-axis and $z$-axis, respectively; $p$ is pressure; and $\eta$ is kinetic viscosity

Energy conservation equation:

$$\rho c_p\left(\frac{\partial t}{\partial \tau} + u\frac{\partial t}{\partial x} + v\frac{\partial t}{\partial y} + w\frac{\partial t}{\partial z}\right) = \lambda\Delta t + \dot{\Phi} \tag{9}$$

where $c_p$ is specific heat at constant pressure.

## 3. Image-based reconstruction of the heat transfer problem

In this study, the heat transfer process of the PFHS is reconstructed based on contour images. The reconstruction procedure is mainly composed of two issues, image regression and image generation which are implemented by supervised learning and unsupervised learning algorithms, respectively.

As shown in Fig. 2, firstly, contour image and corresponding objective function of each iteration during the simulation are collected and slicing technique is used to obtain training samples. Then Convolutional Neural Network (CNN) [33] is employed to construct the mapping from contour images to objective functions and Generative Adversarial Network (GAN) [34] is employed to generate more "pseudo images" in Steps ii and iii, respectively. Thirdly, the objective functions of the generated "pseudo images" by the GAN can be calculated by the trained CNN in Step ii. Finally, the reconstruction is completed by difference for objective functions during the simulation process. In this study, the objective function is the highest temperature.

Nevertheless, experiments in Section 4 of Ref. [31] showed that existing CNNs might be difficult to handle the image regression, and unsatisfied results were also achieved by existing GAN models for mechanical problems. Therefore, a novel CNN architecture with image cutting, Convolution in Convolution (CIC), and an especial GAN model integrated with image compression, Compressed Wasserstein GAN (CWGAN)*, were introduced in Ref. [31].

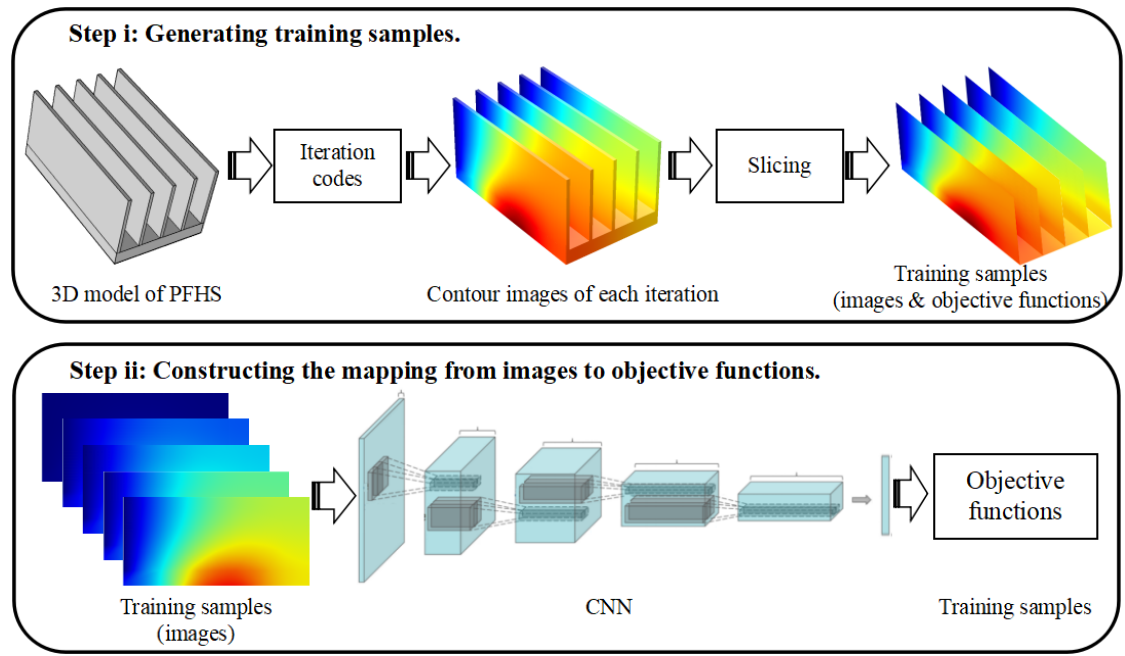

---

* The name of the proposed GAN model has been updated as Compressed WGAN (CWGAN), which was CA-based WGAN (WGAN-CA).

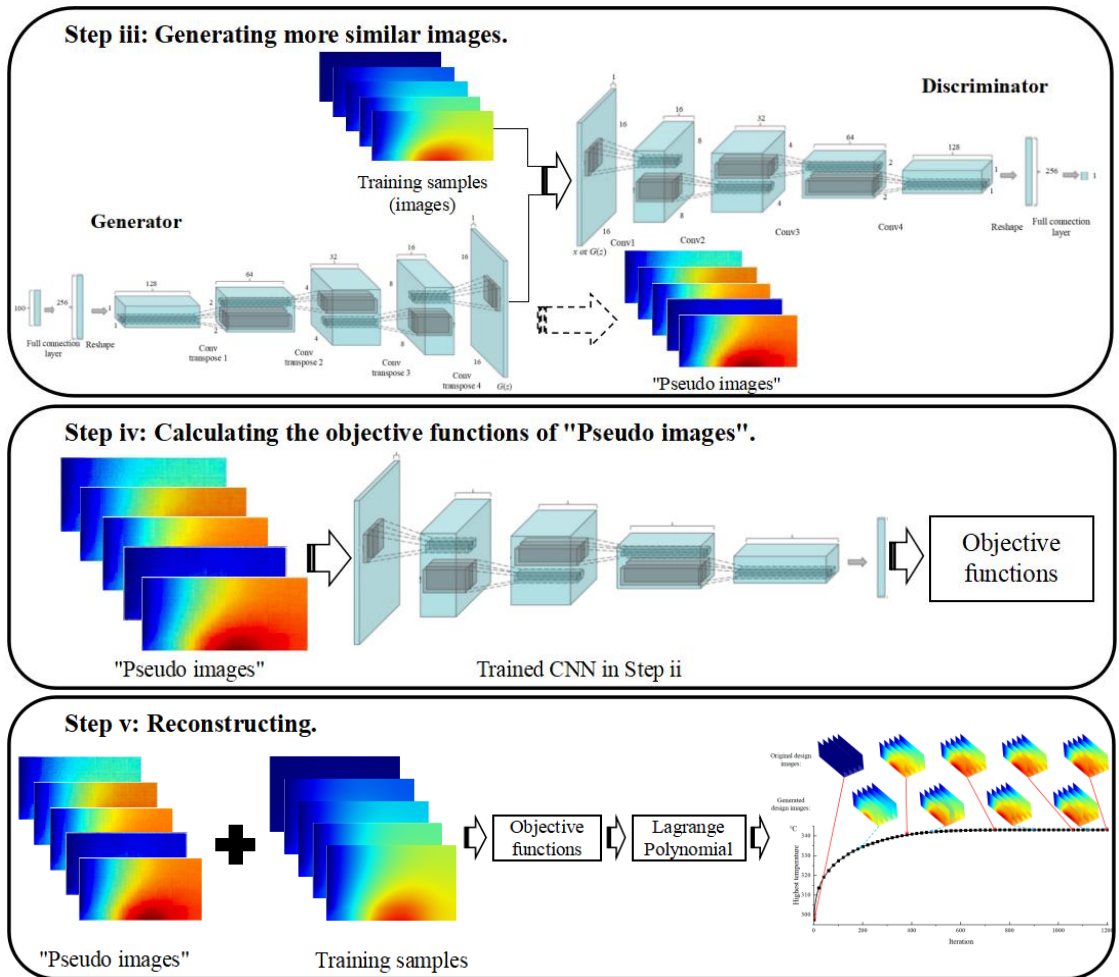

**Fig. 2.** The architecture of the ReConNN model.

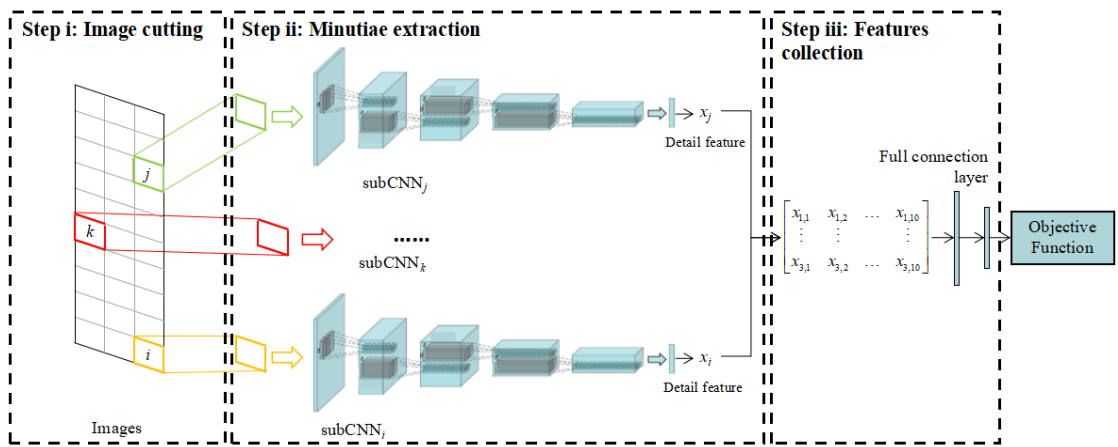

**Fig. 3.** The architecture of the CIC.

As shown in Fig. 3, the CIC can be mainly divided into three steps. In Step i, each image is divided to dozens of subimages by using image cutting technique. Subsequently, the detail features of each subimage are extracted through a separate subCNN. Ultimately, all the features extracted in Step ii are collected to obtain the objective function in the final step. It is well known the details of the simulation are important for designers during optimizations while the popular CNNs might hardly obtain. Therefore, the aim of the CIC is to use separate sub-networks to improve the extraction ability of the detailed features.

The optimization algorithm utilized in the CIC is Adaptive Moment Estimation (Adam) [35] optimizer which is essentially Root Mean Square Prop (RMSProp) [36] with momentum factor. The Adam integrates the advantages of Adaptive Gradient (AdaGrad) [37] and RMSProp, and spends lower computational cost. Furthermore, it performs well for most nonconvex optimization, large data sets, and high-dimensional space. Mathematically, the Adam can be defined as

$$g \leftarrow +\frac{1}{m} \nabla_\theta \sum_i L(f(x_i;\theta), y_i) \tag{10}$$

$$s \leftarrow \rho_1 s + (1-\rho_1) g \tag{11}$$

$$r \leftarrow \rho_2 r + (1-\rho_2) g \odot g \tag{12}$$

$$\hat{s} \leftarrow \frac{s}{1-\rho_1} \tag{13}$$

$$\hat{r} \leftarrow \frac{r}{1-\rho_2} \tag{14}$$

$$\Delta\theta = -\varepsilon \frac{\hat{s}}{\sqrt{\hat{r}}+\delta} \tag{15}$$

$$\theta \leftarrow \theta + \Delta\theta \tag{16}$$

where $L(x)$ is the loss function; $f(x)$ is the prediction of real value $y_i$; $\theta$ is the initial parameter; $x_i$ is a training sample; $m$ is the number of $x_i$; $s$ and $r$ are the first and second moment estimations, respectively; $\rho_1$ and $\rho_2$ are the attenuation coefficients; and $\varepsilon$ is the learning rate. In this study, $\delta=10^{-8}$, $\rho_1=0.9$, and $\rho_2=0.999$.

For the CWGAN, as shown in Fig. 4, in order to improve training accuracy and stability, image compression is introduced through a Convolutional Autoencoder (CA) [38], which is a classical convolutional encoder-decoder architecture. The CA is trained to establish mappings both from the contour images to 16×16×1 gray level images and from 16×16×1 gray level images to the contour images. Through the CA, 16×16×1 gray level images should be the new learning distributions of the WGAN. Therefore, the WGAN-based NN is easier to be trained due to the smaller input sizes.

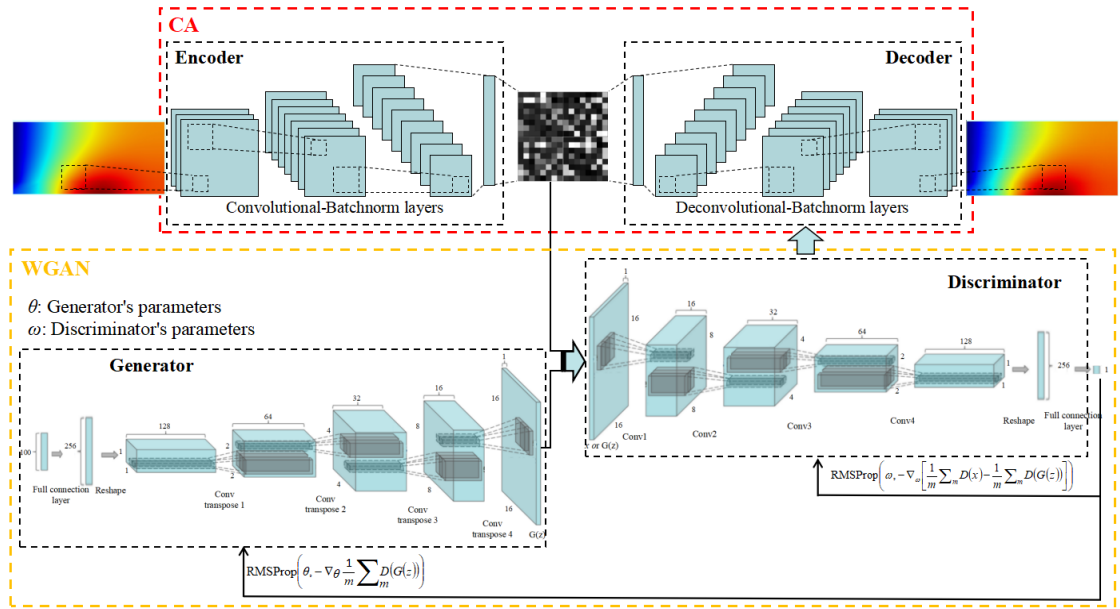

**Fig. 4.** The architecture of the CWGAN.

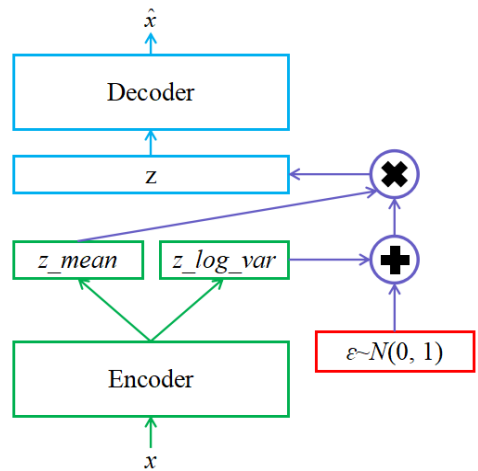

**Fig. 5.** The architecture of the VAE.

In this study, Variational Autoencoder (VAE) [39, 40] is employed as a new compressor. As shown in Fig. 5, compared with the AE, the outputs of the encoder in the VAE have two purposes, one represents the mean of a Gaussian distribution (z_mean), and another means the logarithmic value for the variance of a Gaussian distribution (z_log_var). The input of the decoder can be calculated by

$$z = z\_mean + \varepsilon \cdot \exp\left(\frac{z\_log\_var}{2}\right) \tag{17}$$

where $\varepsilon \sim N(0, 1)$.

The Evidence Lower Bound (ELBO) of the VAE is

$$ELBO = \log p(x) - KL\left[q(z|x) \| p(z|x)\right] \tag{18}$$

where $q$ is the predicted distribution; and $p$ is the real distribution.

To sum up, the detailed architecture of the ReConNN is shown in Table 1. The function of Rectified Linear Unit (ReLU) is calculated by

$$\text{ReLU} = \max(x, 0) \tag{19}$$

The Leaky ReLU (LReLU) used in the discriminator of WGAN can be described as

$$\text{LReLU} = \max(x, \mu x) \tag{20}$$

where $\mu$ is a random value between 0 and 1. In this study, $\mu=0.2$.

Batch normalization allows training to use much higher learning rates and be less careful about initialization [41]. For a layer with $d$-dimensional input $\mathbf{x}=(x_1, \ldots, x_d)$, each dimension is normalized by the batch normalization as

$$y_i = \gamma x_i^- + \beta \tag{21}$$

where

$$x_i^- = \frac{x_i - \mu_\beta}{\sigma_\beta} \tag{22}$$

subject to

$$\mu_\beta = \frac{1}{d}\sum_{i=1}^{d} x_i \tag{23}$$

$$\sigma_\beta^2 = \frac{1}{d}\sum_{i=1}^{d} d(x_i - \mu_\beta)^2 \tag{24}$$

where $\gamma$ and $\beta$ are parameters to be learned; and $\mu_\beta$ and $\sigma_\beta$ are the mean and standard deviation of $x_i$.

The RMSProp optimization used in the WGAN can be expressed as

$$g \leftarrow +\frac{1}{m}\nabla_\theta \sum_i L(f(x_i;\theta), y_i) \tag{25}$$

$$r \leftarrow \rho r + (1-\rho) g \odot g \tag{26}$$

$$\Delta\theta = -\frac{\varepsilon}{\delta+\sqrt{r}} \odot g \tag{27}$$

$$\theta \leftarrow \theta + \Delta\theta \tag{28}$$

**Table 1.** The architecture of the ReConNN.

| | Sub-net | | Network architecture | Activation function | Optimization algorithm |
|---|---|---|---|---|---|
| CIC | | subCNN$_i$ | 3 convolutional-mixed pooling | ReLU | Adam Optimizer |
| | | Full connection layer | 1 convolutional-mixed pooling | ReLU | |
| CWGAN | VAE | Encoder | 2 convolutional-batchnorm-max pooling and 2 full connection | ReLU | Adam Optimizer |
| | | Decoder | 2 full connection, 2 deconvolutional-batchnorm-max pooling | ReLU | |
| | WGAN | Generator | 1 full connection and 2 deconvolutional-batchnorm | ReLU | RMSProp Optimizer |
| | | Discriminator | 2 convolutional-batchnorm and 1 full connection | LReLU | |

## 4. Experiment results

In the previous sections, the CAD model of a 3D-PFHS and the reconstruction method are suggested. In order to evaluate the performance of the reconstruction method for the 3D-PFHS, experiments and analyses are performed in this section.

*4.1. PFHS samples*

As shown in Fig. 6, considering that the 3D-PFHS contains five-plate fins and one baseplate, then six cross sections, from A-A to F-F, are collected.

The reconstruction of the 3D-PFHS can be divided into two issues. Firstly, because of the same geometric parameters and environmental conditions, five-plate fins can be treated as one issue. In this study, 1,211 iterations of the simulation are run. Therefore, there are 6,055 (=1,211×5) samples whose sizes are [850, 420, 3] for the five-plate fins.

Another issue is the baseplate of the PFHS. As mentioned above, 1,211 iterations are run and 1,211 corresponding samples of [850, 630, 3] are collected for the baseplate.

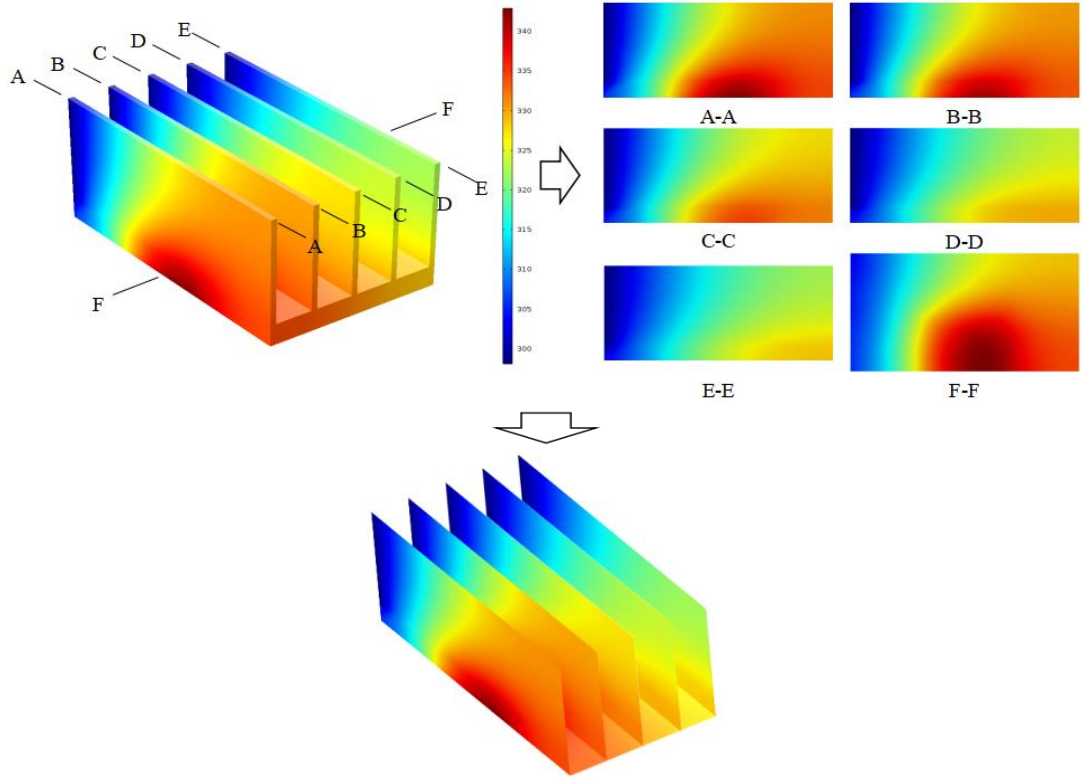

**Fig. 6.** The modeling process of 3D-2D-3D.

*4.2. Training results of the CIC*

In this section, three commonly used performance criteria [42], $R$ Square ($R^2$), Relative Average Absolute Error (RAAE), and Relative Maximum Absolute Error (RMAE), as shown in Table 2, are employed for validating the approximation models by the CIC. Moreover, in order to shown the relative error clearly, the following relative error (*Error*) is used to evaluate the accuracy of the CIC. Among them, $y_i$ is the actual value, $\hat{y}_i$ is the predicted value, STD stands for standard deviation, MSE (Mean Square Error) represents the departure of the metamodel from the real simulation model, and the variance captures the irregular of the problem.

**Table 2** Criteria for performance evaluation.

| Criteria | Expression |
|---|---|
| RAAE | $\sum_{i=1}^{n}|y_i - \hat{y}_i|/(n \cdot \text{STD})$ |
| RMAE | $\max(|y_i - \hat{y}_i|)/\text{STD}$ |
| $R^2$ | $1 - \dfrac{\sum_{i=1}^{n}(y_i - \hat{y}_i)^2}{\sum_{i=1}^{n}(y_i - \bar{y}_i)^2} = 1 - \dfrac{\text{MSE}}{\text{variance}}$ |
| Error | $\dfrac{\|\hat{y} - y\|_2}{\|y\|_2} \times 100\%$ |

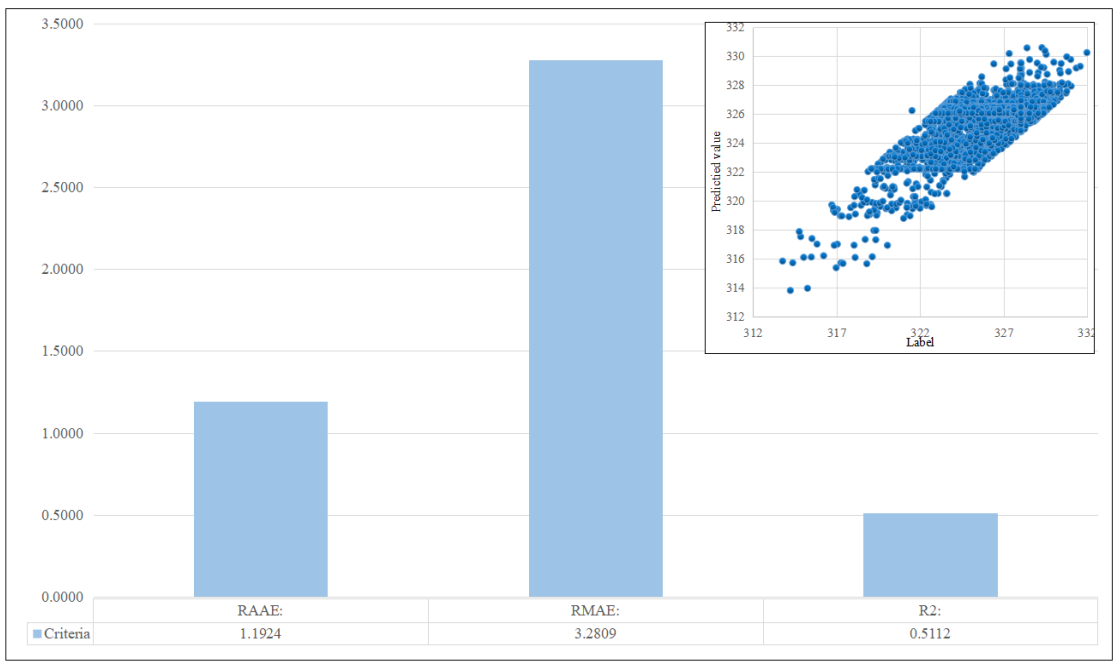

Fig. 7. The training results of the CIC for the five-plate fins.

Subsequently, the samples are utilized to train the CIC for the plate fins and the baseplate. The training results are shown in Figs. 7 - 8. It can be found that $R^2$ for the two issues are more than 0.5, and RAAE and RMAE belong to 1 to 3. Considering the image regressions can be regarded as a thousands-dimensional modeling problem, the criteria are satisfied for the prediction. Moreover, according to the scatter diagram, the distribution of labels* and predicted values is close to the curve of $y=x$, which is a reasonable distribution. In addition, *Error*s correspondingly to the plate fins and the baseplate are 0.5272% and 0.5527%, respectively. They further demonstrate the accuracy of the predicted objective functions.

---

* For NNs, labels denote the real response values of evaluation of samples.

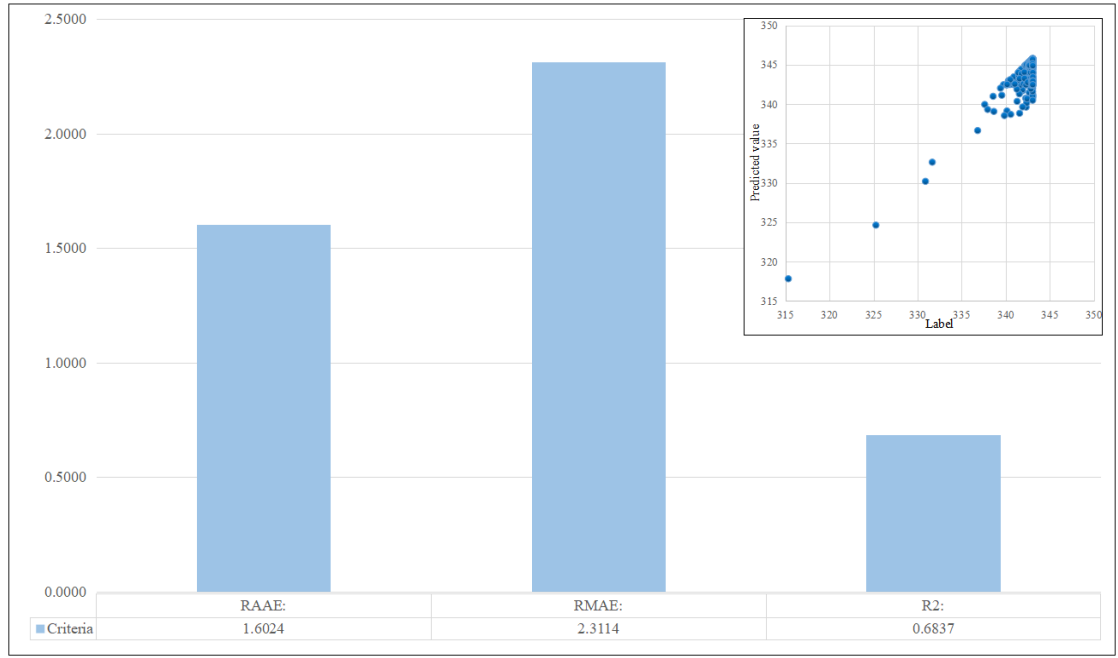

**Fig. 8.** The training results of the CIC for the baseplate.

*4.3. Training results of the CWGAN*

Firstly, the AE is trained. The comparison of the training results between the CA and the VAE are shown in Tables 3 and 4. Compared with the results of the VAE, the images learned by the CA lack convincing details and might be suffered from blurred regions in some cases. For a clearer comparison, two samples are enlarged as shown in Table 5. From the framed areas by green, it can be seen that the result learned by the CA has a clear blurred region. While as shown in the framed areas by red, the CA doesnot learn the sample well. Importantly, the VAE doesnot have such problems and learns the samples well. Therefore, the compressor is replaced by the VAE in this study.

**Table 3.** The training results of the AE for five-plate fins.

| Original images | | | | | | |
|---|---|---|---|---|---|---|
| CA | | | | | | |
| VAE | | | | | | |

**Table 4.** The training results of the AE for the baseplate.

| Original images | 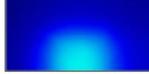 |
|---|---|
| CA | 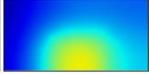 |
| VAE | 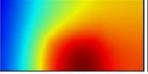 |

**Table 5.** Comparisons of two samples for the fins and the plate, respectively.

| | Original images | CA | VAE |
|---|---|---|---|
| Plate fins | 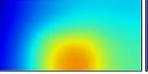 | 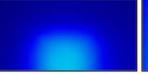 | 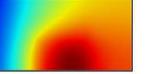 |
| Plate | 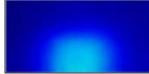 | 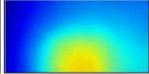 | 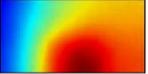 |

Subsequently, the compressed images are employed to train the WGAN. After training, the new 16×16×1 gray level images generated by the generator of the WGAN are reshaped to their original distributions according to the decoder of the AE, which are the purpose of the CWGAN. However, for generative models (e.g. GAN), it is difficult to evaluate their performance. Therefore, a recently proposed numerical assessment approach "inception score" [43] for quantitative evaluation is employed.

$$I = \exp\left(\mathbb{E}_{\mathbf{x}} D_{KL}\left(p(y|\mathbf{x}) \| p(y)\right)\right) \qquad (29)$$

Expand the exponent, then

$$I = \int_x p(y|\mathbf{x})\left[\log p(y|\mathbf{x}) - \log \int_x p(y|\mathbf{x}) dx\right] dx \qquad (30)$$

Nevertheless, infinitesimal calculus cannot be calculated. Therefore, the infinitesimal calculus is solved by using inverse summation operation.

$$I = \frac{1}{N}\sum_i^N p(y|x_i)\left[\log p(y|x_i) - \log \sum_j^N p(y|x_j)\right] \qquad (31)$$

where $\mathbb{E}$ means the operation of mean; **x** denotes sample set; $p(y/\mathbf{x})$ is the softmax output of a trained classifier of the labels; and $p(y)$ is the overall labels distribution of generated samples.

The intuition behind this criterion is that a good model should generate diverse but meaningful images. Therefore, the Kullback-Leibler (KL) divergence between the marginal distribution $p(y)$ and the conditional distribution $p(y|\mathbf{x})$ might be large. However, the score is not calculated directly for all generated images, but instead the generated images are broken up into chunks of size ($N/n_{\text{splits}}$) and the estimator is applied repeatedly on these chunks to compute a mean and standard deviation of the inception score. Generally, $n_{\text{splits}}=10$. The inception scores of the plate fins and baseplate are shown in Table 6.

Table 6. Inception scores of the trained CWGAN.

|  |  | **Plate fins** | **Baseplate** |
|---|---|---|---|
| **CWGAN** | Mean score | 2.820 | 2.592 |
|  | Standard deviation | 1.001e-02 | 3.280e-03 |

Although the inception scores of the CWGAN are likely not very large. Considering that with the process of iteration and converge, the changes between different contour images are slightly. Therefore, theses inception scores might be satisfied and reasonable. Importantly, CWGAN achieves satisfied results not by original but by compressed samples. Characteristics of compressed data are well achieved. It can be inferred that the performance of the CWGAN can be improved through dimensionality reduction.

*4.4. Reconstruction of the heat transfer process*

Each objective function (the highest temperature) of the generated image in Section 4.3 is calculated by the trained CIC in Section 4.2. Then the LI algorithm expressed by Eqs. (32) - (33) is used to complete the reconstruction task.

$$f(x) = \sum_{i=1}^{n} y_i p_i(x), i = 1, 2, \ldots, n \tag{32}$$

subject to

$$p_i = \prod_{i=1, i \neq j}^{n} \frac{x - x_i}{x_j - x_i} = \frac{(x - x_1)(x - x_2)\ldots(x - x_n)}{(x_j - x_1)\ldots(x_j - x_{j-1})(x_j - x_{j+1})\ldots(x_j - x_n)} \tag{33}$$

Considering the heat source is in the baseplate, therefore, the basis of reconstruction is the highest temperature of the baseplate. Finally, the reconstructed 3D heat transfer process is shown in Fig. 9. The images pointed in red are the original process, and the total iterations are 1,211. The images pointed in blue are the generated contour images by CWGAN, and the iteration number of reconstruction is extended to 5,000. Obviously, compared with the original output, about 4000 new samples are generated by the suggested method.

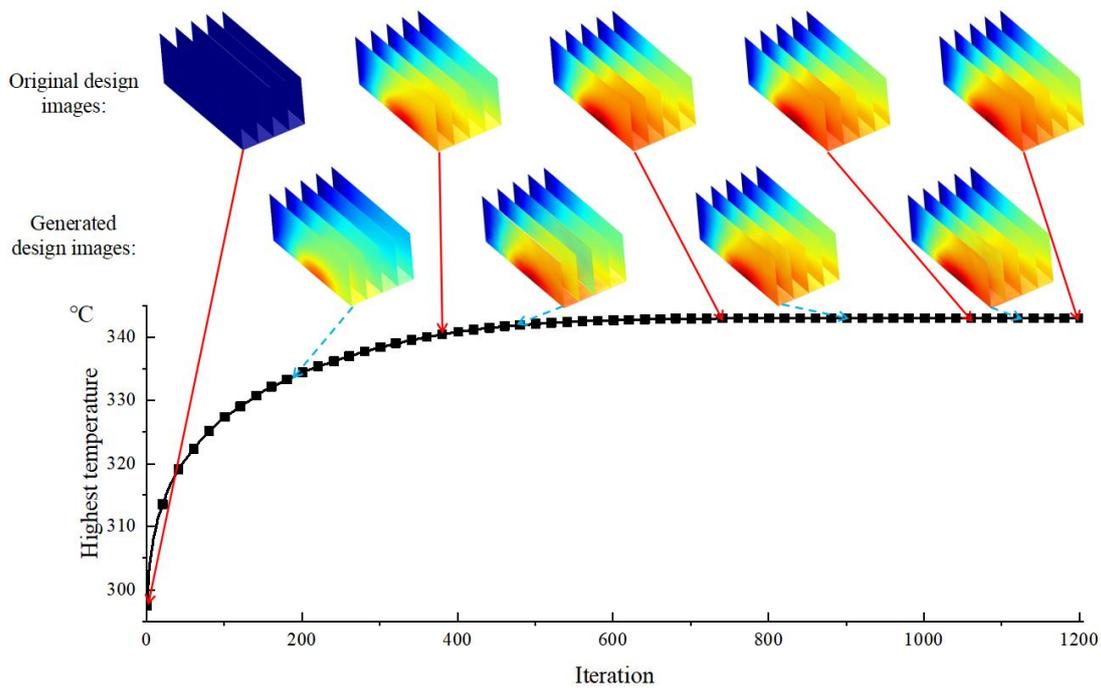

**Fig. 9.** The reconstruction of the heat transfer process of the PFHS.

In order to evaluate the reconstructed results, the generated results by the suggested method are compared with the simulation results directly with 5,000 iterations as shown in Fig. 10. It can be found the results are well matched, and the maximum Relative Error (RE) of the interpolation is 1.728E-2 which is small and satisfied enough.

Compared the original iteration process and LI results. If the heat transfer process is divided into 1,211 iterations according to the original iteration process, then the iteration of the largest change of the highest temperature is $1^{st}$ and $1.3^{rd}$ iteration according to the original iteration process and LI results, respectively. It suggests that more useful data can be applied to modeling in fewer iterations and the accuracy of model based on the new images should be more accurate than the original one.

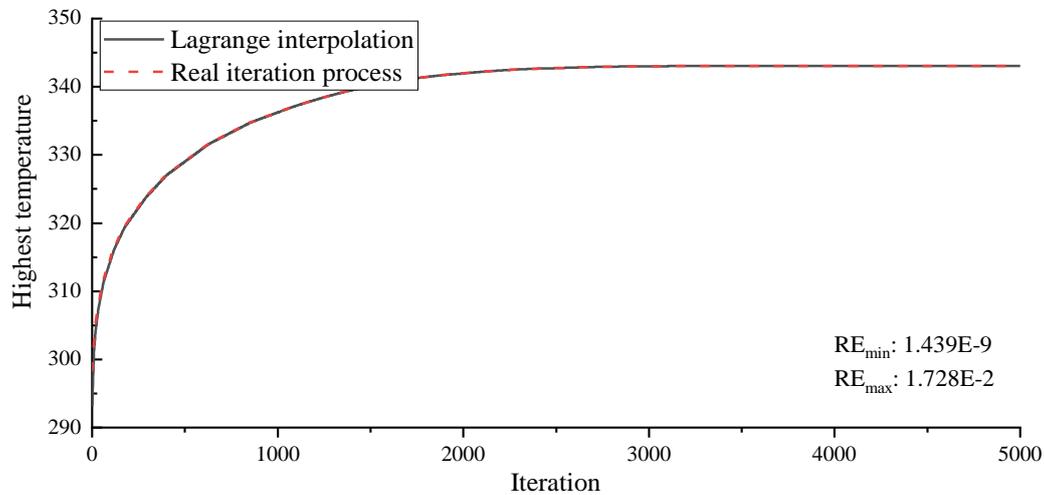

**Fig. 10.** The evaluation for the reconstructed results.

## Conclusions

In this study, the ReConNN model is suggested for reconstruction of a heat transfer processes to construct more objective and more information included models, which can be further applied to researches or some experiments that use high-speed photography to catch some dynamic characteristics and reconstruct the model through these photos. The main contributions of this study can be summarized as follows.

    i. The reconstruction method is implemented from the analysis-based model to the image-based one.

    ii. In order to complete the reconstruction task, a feasibility and efficient ReConNN model is suggested and improved, which is integrated supervised learning and unsupervised learning algorithms.

    iii. A novel reconstruction method, 3D-2D-3D, for 3D models is introduced. It means that the 3D model can be cut into several slices according to its characteristics firstly, and then reconstruction is processed for each slice. Finally, all reconstructed slices are regrouped to a 3D model.

    iv. According to a heat transfer process for a 3D-PFHS, the reconstruction model shows a potential capability to reconstruct a model with more information for

the simulation-based optimization problems.

## Acknowledgments

This work has been supported by Project of the Key Program of National Natural Science Foundation of China under the Grant Numbers 11572120 and 51621004.